\documentclass[10pt]{article}
\usepackage{latexsym,graphicx}
\newcommand{\be}{\begin{equation}}
\newcommand{\ee}{\end{equation}}
\def\n{\noindent}
\catcode `\@=11
\catcode `\@=12
\begin{document}
\begin{center}
\large{\bf { BIANCHI TYPE V UNIVERSES WITH BULK VISCOSITY IN GENERAL RELATIVITY}}\\ 
\vspace{10mm}
\normalsize{ANIRUDH PRADHAN$^a$, VANDANA RAI$^b$ and R. S. SINGH$^c$} \\
\vspace{5mm}
\normalsize{$^a$ Department of Mathematics, Hindu Post-graduate College,
 Zamania-232 331, Ghazipur, India. \\
\normalsize{E-mail : pradhan@iucaa.ernet.in}}\\
\vspace{5mm}
\normalsize{$^{b,c}$ Department of Mathematics, Post-graduate College, 
Ghazipur - 233 001, India \\ 
\normalsize{$^c$E-mail : vandana\_rai005@yahoo.co.in}}
\end{center}
\vspace{10mm}
\begin{abstract} 
Bianchi type V bulk viscous fluid cosmological models are investigated. Using 
a generation technique (Camci {\it et al.}, 2001), it is shown that the Einstein's 
field equations are solvable for any arbitrary cosmic scale function. The viscosity 
coefficient of bulk viscous fluid is assumed to be a power function of mass density. 
Solutions for particular forms of cosmic scale functions are also obtained. Some 
physical and geometric aspects of the models are also discussed.  
\end{abstract}
\smallskip
\n PACS number: {04.20.-q}  \\
\n Keywords: {Bianch type V universe, bulk viscous fluid}
\section{INTRODUCTION}
The study of Bianchi type V cosmological models create more interest as these 
models contain isotropic special cases and permit arbitrary small anisotropy 
levels at some instant of cosmic time. This property makes them suitable as
model of our universe. The homogeneous and isotropic Friedman-Robertson-Walker
(FRW) cosmological models, which are used to describe standard cosmological models,
are particular case of Bianchi type I, V and IX universes, according to whether 
the constant curvature of the physical three-space, $t$ = constant, is zero, 
negative or positive. These models will be interesting to construct cosmological
models of the types which are of class one. Present cosmology is based on the
FRW model which is completely homogeneous and isotropic. This is in agreement with 
observational data about the large scale structure of the universe. However, although 
homogeneous but anisotropic models are more restricted than the inhomogeneous models,
they explain a number of observed phenomena quite satisfactorily. This stimulates the 
research for obtaining exact anisotropic solution for Einstein's field equations 
(EFEs) as a cosmologically accepted physical models for the universe (at least in 
the early stages). Roy and Prasad \cite{ref1}, Bali and Yadav \cite{ref2} have investigated 
Bianchi type V universes which are locally rotationally symmetric and are of 
embedding class one filled with perfect fluid with heat conduction and radiation. 
Bianchi type V cosmological models have been studied by other researchers 
(Farnsworth \cite{ref3}; Maartens and Nel \cite{ref4}; Wainwright et al. \cite{ref5,ref6}; 
Collins \cite{ref7}; Banerjee and Sanyal \cite{ref8}; Meena and Bali \cite{ref9}; 
Pradhan et al. \cite{ref10,ref11}) in different context. 

The role of bulk viscosity in the cosmic evolution, especially as its early 
stages, seems to be significant. The distribution of matter can be satisfactorily 
described by a perfect fluid due to the large scale distribution of galaxies in 
our universe. However, observed physical phenomena such as the large entropy per 
baryon and the remarkable degree of isotropy of the cosmic microwave background radiation, 
suggest analysis of dissipative effects in cosmology. Furthermore, there are
several processes which are expected to give rise to viscous effects. These
are the decoupling of neutrinos during the radiation era and the decoupling
of radiation and matter during the recombination era. Bulk viscosity is
associated with the GUT phase transition and string creation. Misner \cite{ref12}
has studied perfect fluid cosmological models with bulk viscosity and obtained that the 
big-bang singularity may be avoided in the finite past. The
role of viscosity in cosmology has been investigated by Weinberg \cite{ref13}. 
Nightingale \cite{ref14}, Heller and Klimek \cite{ref15} have obtained a viscous 
universes without initial singularity. The model studied by Murphy \cite{ref16} 
possessed an interesting feature in which  big bang type of singularity of infinite 
space-time curvature does not occur to be a finite past. However, the relationship 
assumed by Murphy between the viscosity coefficient and the matter density is not 
acceptable at large density. Thus, we should consider the presence of material 
distribution other than a perfect fluid to obtain a realistic cosmological models 
(see Gr$\o$n \cite{ref17} for a review on cosmological models with bulk viscosity). 
The effect of bulk viscosity on the cosmological evolution has been investigated by 
a number of authors in the framework of general theory of relativity (Kalyani and Singh \cite{ref18}; 
Beesham \cite{ref19}; Singh et al. \cite{ref20}; Bali and Pradhan \cite{ref21}; 
Pradhan et al. \cite{ref22}). This motivates to study cosmological bulk viscous fluid model.
 
In recent years, several authors (Hajj-Boutros \cite{ref23};  Hajj-Boutros and Sfeila \cite{ref24}; 
Ram \cite{ref25}; Mazumder \cite{ref26}; and Pradhan and Kumar \cite{ref27}) have investigated 
the solutions of EFEs for homogeneous but anisotropic models by using some different generation 
techniques. Bianchi spaces $I$-$IX$ are useful tools in constructing models of spatially 
homogeneous cosmologies (Ellis and MacCallum \cite{ref28}; Ryan and Shepley \cite{ref29}). 
From these models, homogeneous Bianchi type V universes are the natural generalization of the 
open FRW model which eventually isotropize. Recently Camci et al. \cite{ref30}
derived a new technique for generating exact solutions of EFEs with perfect fluid 
for Bianchi type V space-time. Very recently Pradhan et al. \cite{ref31} have obtained 
Bianchi type V perfect fluid cosmological models with time dependent $\Lambda$-term.

In this paper, in what follows, we will discuss Bianchi type V cosmological models obtained 
by augmenting the energy-momentum tensor of a bulk viscous, and later generalize the solutions 
(Camci et al. \cite{ref30}; Ram \cite{ref25}; Pradhan and Kumar \cite{ref27}). This paper is 
organized as follows: The field equations and the generation technique are presented in Section $2$. 
We relate three of the metric variables by solving the off-diagonal component of EFEs, and find a 
second integral which is used to relate the remaining two metric variables. In Section 3, for the 
particular form of each metric variables, some solutions are presented separately and solutions of 
Camci et al. \cite{ref30} and Ram \cite{ref25} are shown to be particular cases of 
our solutions. Kinematical and dynamical properties of all solutions are also studied in 
this section. In Section $4$, we give the concluding remarks. 

\section{FIELD EQUATIONS AND GENERATION TECHNIQUE} 
In this section, we review the solutions obtained by Camci et al. \cite{ref30}.

We consider the space-time metric of the spatially homogeneous Bianchi type V of
the form
\begin{equation}
\label{eq1}
ds^{2} = dt^{2} - A^{2}(t) dx^{2} - e^{2\alpha x} \left[B^{2}(t) dy^{2} + C^{2}(t)dz^{2}\right],
\end{equation}
where $\alpha$ is a constant. 

The stress energy-momentum tensor in presence of bulk stress is given by
\begin{equation}
\label{eq2}
T_{ij} = (\bar{p} + \rho)u_{i}u_{j} - \bar{p} g_{ij},
\end{equation}
where 
\begin{equation}
\label{eq3}
\bar{p} = p + \xi u^{i}_{;i}
\end{equation}
Here, $\rho$, $p$, $\bar{p}$, $\xi$ and $u$ are, respectively, the energy density, isotropic 
pressure, effective pressure, bulk viscous coefficient and the fluid four-velocity vector of 
distribution such that $u^{i} u_{i} = 1$. In general, $\xi$ is a function of time. The 
semicolon $(;)$ indicates covariant differentiation. We choose the coordinates to be comoving, 
so that $u^{1} = 0 = u^{2} = u^{3}$, $u^{4} = 1$. 

For the energy momentum tensor (\ref{eq2}) and Bianchi type V space-time (\ref{eq1}), 
Einstein's field equations
\begin{equation}
\label{eq4}
R_{ij} - \frac{1}{2} R g_{ij} = - 8\pi T_{ij}
\end{equation}
yield the following five independent equations
\begin{equation}
\label{eq5}
\frac{A_{44}}{A} + \frac{B_{44}}{B} + \frac{A_{4}B_{4}}{AB} - \frac{\alpha^{2}}
{A^{2}} = - 8\pi \bar{p},
\end{equation}
\begin{equation}
\label{eq6}
\frac{A_{44}}{A} + \frac{C_{44}}{C} + \frac{A_{4}C_{4}}{AC} - \frac{\alpha^{2}}
{A^{2}} = - 8\pi \bar{p},
\end{equation}
\begin{equation}
\label{eq7}
\frac{B_{44}}{B} + \frac{C_{44}}{C} + \frac{B_{4}C_{4}}{BC} - \frac{\alpha^{2}}
{A^{2}} = - 8\pi \bar{p},
\end{equation}
\begin{equation}
\label{eq8}
\frac{A_{4}B_{4}}{AB} + \frac{A_{4}C_{4}}{AC} + \frac{B_{4}C_{4}}{BC} - 
\frac{3\alpha^{2}}{A^{2}} = 8\pi \rho,
\end{equation}
\begin{equation}
\label{eq9}
\frac{2A_{4}}{A} - \frac{B_{4}}{B} - \frac{C_{4}}{C} = 0.
\end{equation}
Here and in what follows the suffix $4$ by the symbols $A$, $B$, $C$ and $\rho$ denote 
differentiation with respect to $t$. The Bianchi identity ($T^{ij}_{;j} = 0$) takes the 
form
\begin{equation}
\label{eq10}
{\rho}_{4} + (\rho + \bar{p}) \theta = 0.
\end{equation}
It is worth noting here that our approach suffers from a lack of Lagrangian 
approach. There is no known way to present a consistent Lagrangian model
satisfying the necessary conditions discussed in this paper. 

The physical quantities expansion scalar $\theta$ and shear scalar $\sigma^{2}$
have the following expressions:
\begin{equation}
\label{eq11}
\theta = u^{i}_{;i} = \frac{A_{4}}{A} + \frac{B_{4}}{B} + \frac{C_{4}}{C}
\end{equation} 
\begin{equation}
\label{eq12}
\sigma^{2} = \frac{1}{2}\sigma_{ij}\sigma^{ij} = \frac{1}{3}\left[\theta^{2} - 
\frac{A_{4}B_{4}}{AB} - \frac{A_{4}C_{4}}{AC} - \frac{B_{4}C_{4}}{BC}\right].
\end{equation} 
Integrating Eq. (\ref{eq9}) and absorbing the integration constant into $B$ or
$C$, we obtain
\begin{equation}
\label{eq13}
A^{2} = BC
\end{equation} 
without any loss of generality. Thus, elimination of $\bar{p}$ from
Eqs. (\ref{eq5}) - (\ref{eq7}) gives the condition of isotropy of pressures
\begin{equation}
\label{eq14}
2\frac{B_{44}}{B} + \left(\frac{B_{4}}{B}\right)^{2} = 2\frac{C_{44}}{C} + 
\left(\frac{C_{4}}{C}\right)^{2},
\end{equation} 
which on integration yields
\begin{equation}
\label{eq15}
\frac{B_{4}}{B} - \frac{C_{4}}{C} = \frac{k}{(BC)^{3/2}},
\end{equation}
where $k$ is a constant of integration. Hence for the metric function $B$ or 
$C$ from the above first order differential Eq. (\ref{eq15}), some scale 
transformations permit us to obtain new metric function $B$ or $C$. \\

Firstly, under the scale transformation $dt = B^{1/2}d\tau$, Eq. (\ref{eq15}) 
takes the form     
\begin{equation}
\label{eq16}
C B_{\tau} - B C_{\tau} = k C^{-1/2},
\end{equation}
where subscript represents derivative with respect to $\tau$. Considering Eq. 
(\ref{eq16}) as a linear differential equation for $B$, where $C$ is an arbitrary
function, we obtain
\begin{equation}
\label{eq17}
(i) B = k_{1} C + k C \int{\frac{d\tau}{C^{5/2}}},
\end{equation}
where $k_{1}$ is an integrating constant. Similarly, using the transformations
$dt = B^{3/2}d\tilde{\tau}$, $dt = C^{1/2}dT$, and $dt = C^{3/2} d\tilde{T}$
in Eq. (\ref{eq15}), after some algebra we obtain respectively 
\begin{equation}
\label{eq18}
(ii) B(\tilde{\tau}; k_{2}, k) = k_{2} C ~ \exp{\left(k\int{\frac{d\tilde{\tau}}
{C^{3/2}}}\right)}, 
\end{equation}
\begin{equation}
\label{eq19}
(iii) C(T; k_{3}, k) = k_{3} B - k B \int{\frac{dT}{B^{5/2}}},
\end{equation}
and
\begin{equation}
\label{eq20}
(iv) C(\tilde{T}; k_{4}, k) = k_{4} B ~ \exp{\left(k\int{\frac{d\tilde{T}}
{B^{3/2}}}\right)}, 
\end{equation}
where $k_{2}$, $k_{3}$ and $k_{4}$ are constants of integration. Thus choosing 
any given function $B$ or $C$ in cases (i), (ii), (iii) and (iv), one can obtain 
$B$ or $C$ and hence $A$ from (\ref{eq13}).

\section{GENERATION OF NEW SOLUTIONS} 

We consider the following four cases:  

\subsection{Case (I): Let $C = \tau^{n}$, ($n$ is a real number satisfying
$n \ne \frac{2}{5}$).}  
In this case, Eq. (\ref{eq17}) gives
\begin{equation}
\label{eq21}
B = k_{1}\tau^{n} + \frac{2k}{2 - 5n}\tau^{1 - 3n/2}
\end{equation}
and then from (\ref{eq13}), we obtain
\begin{equation}
\label{eq22}
A^{2} = k_{1}\tau^{2n} + \frac{2k}{2-5n}\tau^{1 - n/2}.
\end{equation}
Hence the metric (\ref{eq1}) reduces to the new form
\begin{equation}
\label{eq23}
ds^{2} = \left(k_{1}\tau^{n} + 2\ell \tau^{\ell_{1}}\right)[d\tau^{2} - \tau^{n} 
dx^{2}] - e^{2\alpha x}\left[\left(k_{1}\tau^{n} + 2\ell \tau^{\ell_{1}}\right)^
{2} dy^{2} + \tau^{2n} dz^{2}\right],
\end{equation}
where
\[
\ell = \frac{k}{2 - 5n} ~ and ~ \ell_{1} = 1 - \frac{3n}{2}.
\]
For this derived model (\ref{eq23}), the effective pressure and energy density are given by
\[
8\pi \bar{p} = D_{1}^{-3} \biggl[-2k^{2}_{1} n(n - 1)\tau^{2n -2} - k_{1}\ell n(10 - 13n)
\tau^{-(\ell_{1} + 2n)} - 
\] 
\begin{equation}
\label{eq24}
\frac{\ell^{2}(4 + 4n - 11n^{2})}{2}\tau^{-3n}\biggr] + \alpha^{2}\tau^{-n}D_{1}^{-1},
\end{equation}
\[
8\pi \rho = D_{1}^{-3}\biggl[3k^{2}_{1} n^{2}\tau^{2n -2} + 3k_{1}\ell n(2 - n)
\tau^{-(\ell_{1} + 2n)} + 
\] 
\begin{equation}
\label{eq25}
 \frac{\ell^{2}(4 + 4n - 11n^{2})}{2}\tau^{-3n}\biggr]  - 3\alpha^{2}\tau^{-n} D_{1}^{-1},
\end{equation}
where
$$ D_{1} = \left(k_{1}\tau^{n} + 2\ell \tau^{\ell_{1}}\right). $$

The metric (\ref{eq23}) is a four-parameter family of solutions to EFEs with a
perfect fluid. Using the scale transformation $dt = B^{\frac{1}{2}} d\tau$ in Eqs. (\ref{eq11})
and (\ref{eq12}) for this case, the scalar expansion $\theta$ and the shear $\sigma$ 
have the expressions:
\begin{equation}
\label{eq26}
\theta = 3\left[k_{1} n \tau^{n - 1} + \frac{\ell(2 - n)}{2} \tau^{-3n/2}\right]
D_{1}^{-3/2}
\end{equation} 
\begin{equation}
\label{eq27}
\sigma = \frac{1}{2} k \tau^{-3n/2}D_{1}^{-3/2}
\end{equation}
Eqs. (\ref{eq26}) and (\ref{eq27}) lead to
\begin{equation}
\label{eq28}
\frac{\sigma}{\theta} = \frac{k}{6}\left[k_{1} n \tau^{n - \ell_{1}} + \frac{\ell(2 - n)}
{2}\right]^{-1}
\end{equation}
Thus, for given $\xi(t)$, we can solve the system for the physical
quantities. In most of investigations involving bulk viscosity it is assumed 
to be a simple power function of the energy density (Pavon et al. \cite{ref32}; 
Maartens \cite{ref33}; Zimdahl \cite{ref34}; Santos et al. \cite{ref35})
\begin{equation}
\label{eq29} 
\xi(t) = \xi_{0} \rho^{w},
\end{equation}
where $\xi_{0}$ and $w$ are real constants. For small density, $w$ may even be 
equal to unity as used in Murphy's work \cite{ref16} for simplicity. If $w = 1$, Eq.
(\ref{eq29}) may correspond to a radiative fluid \cite{ref13}. Near the 
big bang, $0 \leq w \leq \frac{1}{2}$ is a more appropriate assumption \cite{ref36} to 
obtain realistic models.

For simplicity and realistic models of physical importance, we consider the following 
two cases ($w = 0, 1$):  
\subsubsection{Model I: ~ ~ Solution for $w = 0$}
When $w = 0$, Eq. (\ref{eq29}) reduces to $\xi = \xi_{0}$ = constant. Hence in this case 
Eq. (\ref{eq24}), with the use of (\ref{eq26}), leads to
\[
8\pi p = 3\xi_{0}\Big[k_{1} n \tau^{n - 1} + \frac{\ell(n _ 1)}{2}\tau^{-\frac{3n}{2}}\Big]D_{1}^{-3/2} 
+ D_{1}^{-3} \biggl[-2k^{2}_{1} n(n - 1)\tau^{2n -2} -
\tau^{-(\ell_{1} + 2n)} - 
\] 
\begin{equation}
\label{eq30}
k_{1}\ell n(10 - 13n)\tau^{-(\ell_{1} + 2n)} - \frac{\ell^{2}(4 + 4n - 11n^{2})}{2}\tau^{-3n}\biggr] 
+ \alpha^{2}\tau^{-n}D_{1}^{-1}.
\end{equation}
\subsubsection{Model II: ~ ~ Solution for $w = 1$}
When $w = 1$, Eq. (\ref{eq28}) reduces to $\xi = \xi_{0}\rho$. Hence in this case 
Eq. (\ref{eq24}), with the use of (\ref{eq26}), leads to
\[
8\pi p = 3\xi_{0}\Big[k_{1}n\tau^{n - 1} + \frac{\ell(2 - n)}{2} \tau^{-\frac{3}{2}}\Big] D_{1}^{-3/2}\times
\]
\[
D_{1}^{-3}\biggl[3k^{2}_{1} n^{2}\tau^{2n -2} + 3k_{1}\ell n(2 - n)
\tau^{-(\ell_{1} + 2n)} + 
\] 
\[
 \frac{\ell^{2}(4 + 4n - 11n^{2})}{2}\tau^{-3n}\biggr]  - 3\alpha^{2}\tau^{-n} D_{1}^{-1} -
\]
\[
D_{1}^{-3} \biggl[-2k^{2}_{1} n(n - 1)\tau^{2n -2} - k_{1}\ell n(10 - 13n)
\tau^{-(\ell_{1} + 2n)} - 
\] 
\begin{equation}
\label{eq31}
\frac{\ell^{2}(4 + 4n - 11n^{2})}{2}\tau^{-3n}\biggr] + \alpha^{2}\tau^{-n}D_{1}^{-1},
\end{equation}
From Eq. (\ref{eq25}), we observe that at the time of early universe the 
the energy density $\rho(t)$ decreases with time increases . 

Now, we consider three subcases for the parameters $n$, $k$, $k_{1}$ whether 
zero or not. \\

In subcase $n = 0$, after a suitable inverse time transformation, 
we find that 
\begin{equation}
\label{eq32}
ds^{2} = dt^{2} - K_{1}(t + t_{0})^{2/3}dx^{2} - e^{2\alpha x}\left[K_{1}(t + t_{0})^
{4/3}dy^{2} + dz^{2}\right],
\end{equation}
where $t_{0}$ is a constant of integration and $K_{1} = (3k/2)^{2/3}$. The expressions 
$p$, $\rho$, $\theta$ and $\sigma$ for this model are not given here, since it is 
observed that the physical properties of this one are same as that of the solution 
(\ref{eq24}) of Ram \cite{ref25}. \\

In subcase $k =0$, after inverse time transformation and rescaling,
the metric (\ref{eq23}) reduces to
\begin{equation}
\label{eq33}
ds^{2} = dt^{2} - K_{2}(t + t_{1})^{\frac{4n}{n + 2}}\left[dx^{2} + e^{2\alpha x}
(dy^{2} + dz^{2})\right],
\end{equation}  
where $t_{1}$ is a constant of integration and $K_{2} = \left(\frac{n + 2}{2}\right)
^{\frac{4n}{n + 2}}$. For this solution, when $n = 1$ and $\alpha = 0$, we obtain
Einstein and de Sitter \cite{ref37} dust filled universe. For $K_{2} = 1$, $t_{1} 
= 0$ and $ n = \frac{2m}{(2 - m)}$, where $m$ is a parameter in Ram's paper \cite{ref25}, 
the solution (\ref{eq23}) reduces to the metric (\ref{eq14}) of Ram \cite{ref25}. In later case, 
if also $\alpha = 0$, then we get the Minkowski 
space-time. \\

Now, in subcase $k_{1} = 0$, after some algebra the metric (\ref{eq23})
takes the form
\begin{equation}
\label{eq34}
ds^{2} = dt^{2} - 2\ell K_{3} (t + t_{2})^{2/3}\left[dx^{2} + e^{2\alpha x} \left(
a t^{m_{1}} dy^{2} + a^{-1}t^{-m_{1}} dz^{2}\right)\right],
\end{equation} 
where $t_{2}$ is a constant, $at^{m_{1}} = 2\ell K^{\frac{2 - 5n}{2 - n}}_{3} (t + t_{2})
^{\frac{2(2 - 5n)}{3(2 - n)}}$ and $ K_{3} = \left[\frac{3(2 - n)}{4\sqrt{2 \ell}}
\right]^{2/3}$. \\
For $t_{2} = 0$, $k = \frac{2}{3}$ and $n = 0$ from (\ref{eq34}), we obtain that the 
solution (\ref{eq24}) of Ram \cite{ref25}.\\  


{\bf {SOME PHYSICAL ASPECTS OF MODEL} :} \\

The model (\ref{eq23}) has barrel singularity at $\tau = \tau_{0}$ given by
\[
\tau_{0} = \left[\frac{k_{1}(5n - 2)}{2k}\right]^{\frac{2}{(2 - 5n)}},
\]
which corresponds to $t = 0$. For $n \ne 2/5$ from (\ref{eq23}), it is observed that 
at the singularity state $\tau = \tau_{0}$, $p$, $\rho$, $\theta$ and 
$\sigma$ are infinitely large. At $t \to \infty$, which corresponds to $\tau \to 
\infty$ for $n < 2/5$ and $k> 0$, or $\tau \to 0$ for $n > 2/5$ and $k < 0$, $p$, $\rho$, 
$\Lambda$, $\theta$ and $\sigma$ vanish. Therefore, for $n \ne 2/5$, the solution
(\ref{eq23}) represents an anisotropic universe exploding from $\tau = \tau_{0}$, i.e.
$t = 0$, which expands for $0 < t < \infty$. We also find that the ratio $\sigma / \theta$
tends to a finite limit as $t \to \infty$, which means that the shear scalar does not 
tend to zero faster than the expansion. Hence the model does not approach isotropy for
large values of $t$. 

In subcase $k = 0$, the ratio (\ref{eq28}) tends to zero, then the model
approaches isotropy i.e. shear scalar $\sigma$ goes to zero. For the model (\ref{eq33}),
$p$ and $\rho$ tends to zero as $t \to \infty$; the model would give an essentially empty
universe at large time. The dominant energy condition given by Hawking and Ellis \cite{ref38} 
requires that 
\begin{equation}
\label{eq35}
\rho + p \geq 0, ~ ~ \rho + 3p \geq 0
\end{equation} 
Thus, we find for the model (\ref{eq34}) that $n(2 - n) \geq 0$. Hence for the values
$0 \leq n \leq 2$, the universe (\ref{eq34}) satisfies the strong energy condition i.e.
$\rho + 3p \geq 0$. Also this model is sheer-free and expanding. \\ 

In subcase $k_{1} = 0$, for $n \neq 2/5$, $2$, it is observed from relations
(\ref{eq25}) - (\ref{eq27}), (\ref{eq30}) and (\ref{eq31}) that $\rho$, $\theta$, 
$\sigma$ and $p$ are infinitely large at the singularity state $t = - t_{2}$. When 
$t \to \infty$, these quantities vanish. We also find that the ratio $\sigma/ \theta$ is 
a constant. This shows that the cosmological  model (\ref{eq34}) does not approach 
isotropy for large value of $t$. In this model the dominant energy conditions (\ref{eq35}) 
are then verified for $6 - 5n - 25n^{2} \geq 0$. Since $n \ne 2/5$, the model (\ref{eq34}) 
satisfies the strong energy condition for $-3/5 \leq n \leq 2/5$. \\

In each of subcases, all the obtained solutions (\ref{eq32}), (\ref{eq33}) and 
(\ref{eq34}) satisfy the Bianchi identity given in Eq. (\ref{eq10}).  

\subsection{Case (II): Let $C = \tilde{\tau}^{n}$, ($n$ is a real number 
satisfying $n \ne 2/3$).} 
In this case Eq. (\ref{eq18}) gives
\begin{equation}
\label{eq36}
B = k_{2} \tilde{\tau}^{n} \exp{\left(M\tilde{\tau}^{\ell_{1}}\right)}
\end{equation} 
and from (\ref{eq13}), we obtain
\begin{equation}
\label{eq37}
A^{2} = k_{2} \tilde{\tau}^{2n} \exp{\left(M\tilde{\tau}^{\ell_{1}}\right)},
\end{equation} 
where $M = \frac{k}{\ell_{1}}$. Hence the metric (\ref{eq1}) reduces to the form
\[
ds^{2} = \tilde{\tau}^{4(1 - \ell_{1})/3}\biggl[\tilde{\tau}^{2(1 - \ell_{1})/3} 
e^{3M\tilde{\tau}^{\ell_{1}}} d\tilde{\tau}^{2} - e^{M\tilde{\tau}^{\ell_{1}}} dx^{2}
\]  
\begin{equation}
\label{eq38}
- e^{2\alpha x}\left(e^{2M\tilde{\tau}^{\ell_{1}}} dy^{2} + dz^{2}\right)\biggr],
\end{equation}
where the constant $k_{2}$ is taken, without any loss of generality, equal to $1$.
This metric is a three-parameter family of solutions to EFEs with a perfect fluid.\\

For the above model, the distribution of matter and nonzero kinematical parameters
are obtained as  
\[
8\pi \bar{p} = 2n\tilde{\tau}^{2(\ell_{1} - 2)} + 3nk\tilde{\tau}^{3\ell_{1} - 4}
\]
\begin{equation}
\label{eq39}
+ \frac{k^{2}}{2}\tilde{\tau}^{4(\ell_{1} - 1)} + \alpha^{2}\tilde{\tau}^
{4(\ell_{1} - 1)/3} e^{-3M\tilde{\tau}^{\ell_{1}}},
\end{equation} 
\[
8\pi \rho = 3n^{2}\tilde{\tau}^{2(\ell_{1} - 2)} + 3nk\tilde{\tau}^
{3\ell_{1} - 4}
\]
\begin{equation}
\label{eq40}
+ \frac{k^{2}}{2}\tilde{\tau}^{4(\ell_{1} - 1)} - 3\alpha^{2}\tilde{\tau}^
{4(\ell_{1} - 1)/3} e^{-3M\tilde{\tau}^{\ell_{1}}}.
\end{equation} 
The scalar of expansion $\theta$ and the shear $\sigma$ are obtained as
\begin{equation}
\label{eq41}
\theta = 3\left[n\tilde{\tau}^{\ell_{1} - 2} + \frac{k}{2}\tilde{\tau}^{2(\ell_{1} 
- 1)}\right]e^{-3M\tilde{\tau}^{\ell_{1}}},
\end{equation} 
\begin{equation}
\label{eq42}
\sigma = \frac{k}{2}\tilde{\tau}^{2(\ell_{1} - 1)}e^{-3M\tilde{\tau}^{\ell_{1}}},
\end{equation} 
From Eqs. (\ref{eq41}) and (\ref{eq42}), we have
\begin{equation}
\label{eq43}
\frac{\sigma}{\theta} = \frac{k}{6\left(n \tilde{\tau}^{-\ell_{1}} + \frac{k}{2}\right)}.
\end{equation}
For simplicity and realistic models of physical importance, we consider the following 
two cases ($w = 0, 1$):  

\subsubsection{Model I: ~ ~ Solution for $w = 0$}
When $w = 0$, Eq. (\ref{eq29}) reduces to $\xi = \xi_{0}$ = constant. Hence in this case 
Eq. (\ref{eq39}), with the use of (\ref{eq41}), leads to
\begin{equation}
\label{eq44}
8\pi p = 24\pi \xi_{0} D_{2} + 2n\tilde{\tau}^{2(\ell_{1} - 2)} + 3nk\tilde{\tau}^{3\ell_{1} - 4}
+ \frac{k^{2}}{2}\tilde{\tau}^{4(\ell_{1} - 1)} + D_{3}
\end{equation}
where
$$D_{2} = n \tilde{\tau}^{(\ell_{1} - 2)} + \frac{1}{2}k \tilde{\tau}^{2(\ell_{1} - 1)},$$

$$D_{3} = \alpha^{2} \tilde{\tau}^{\frac{4}{3}(\ell_{1} - 1)}e^{-3M \tilde{\tau}^{\ell_{1}}}.$$ 

\subsubsection{Model II: ~ ~ Solution for $w = 1$}
When $w = 1$, Eq. (\ref{eq29}) reduces to $\xi = \xi_{0}\rho$ . Hence in this case 
Eq. (\ref{eq39}), with the use of (\ref{eq41}), leads to
\[
8\pi p = n \tilde{\tau}^{(\ell_{1} - 2)}(3\xi_{0}nD_{2} + 2) + 3nk \tilde{\tau}^{3\ell_{1} - 4)}
(\xi_{0}D_{2} + 1)
\]
\begin{equation}
\label{eq45}
+ \frac{k^{2}}{2}\tilde{\tau}^{4(\ell_{1} - 1)}(D_{2} + 1) - D_{3}(3D_{2} + 1) 
\end{equation}

From Eq. (\ref{eq40}), we observe that the energy density $\rho(t)$ 
decreases with time increases. Here we find energy density always positive.

In subcase $\xi_{0} = 0$, metric (\ref{eq38}) with expressions $p$, $\rho$, $\theta$ 
and $\sigma$ for this model are same as that of solution (\ref{eq27}) of Camci 
et al. \cite{ref30} . 

In sub-case $\xi_{0} = 0$, $\ell_{1} = 1$ (i.e. $n = 0$), we find a similar solution to
(\ref{eq32}), and hence this subclass is omitted. For $k =0$, the ratio (\ref{eq43}) 
is zero and hence there is no anisotropy. \\

After a suitable coordinate transformation, the metric (\ref{eq38}) can be written as
\begin{equation}
\label{eq46}
ds^{2} = dt^{2} - K_{4}(t + t_{3})^{2\ell_{1}}\left[dx^{2} + e^{2\alpha x}(dy^{2} 
+ dz^{2})\right],
\end{equation} 
where $t_{3}$ is a constant and $K_{4} = \left[\frac{2}{(2 - 3M_{1})}\right]^{2M_{1}}$, 
$M_{1} = \frac{2n}{2 + 3n} \ne \frac{2}{3}$, where $M_{1}$ is a new parameter. When $M_{1} = 0$ 
and $\ell_{1} = 0$, from (\ref{eq46}), we get the solution (\ref{eq12}) of Ram \cite{ref25}.\\

{\bf {SOME PHYSICAL ASPECTS OF THE MODEL}:} 

The models have singularity at $\tilde{\tau} \to - \infty$ for $\ell_{1} > 0$
or $\tilde{\tau} \to 0$ for $\ell_{1} < 0$, which corresponds to $t \to 0$.
It is a point type singularity for $\ell_{1} > 0$ whereas it is a cigar or a barrel 
singularity according as $\ell_{1} < 0$. At $t \to \infty$, which correspond to
$\tilde{\tau} \to \infty$ for $\ell_{1} > 0$ or $\tilde{\tau} \to 0$ for 
$\ell_{1} < 0$, from Eqs. (\ref{eq39}) - (\ref{eq42}), we obtain that for
$\ell_{1} > 0$, $p,\rho \to 0$, and $\sigma, \theta \to 0$ ($k > 0$), - $\infty$
($k < 0$); for  $\ell_{1} < 0$, similar the above ones. Then, clearly, for a
realistic universe, it must be fulfill as $\tilde{\tau} \to - \infty$, $n$ and $k$ 
are positive and $\ell_{1}$ is an odd positive number; as $\tilde{\tau} \to 0$,
$k$ is positive, and $\ell_{1}$ an even negative number. Also, since $lim_{\tilde{\tau} \to
\infty} \frac{\sigma}{\theta} \ne 0$, therefore these models do not approach isotropy
for large values of $\tilde{\tau}$. \\

In subcase $k = 0$ for the metric (\ref{eq46}), the effective pressure and density are given by
\begin{equation}
\label{eq47}
8\pi \bar{p}  = \frac{\ell_{1}(2 - 3\ell_{1})}{(t + t_{3})^{2}} + \frac{\alpha^{2}}
{K_{4}(t + t_{3})^{2\ell_{1}}},
\end{equation} 
\begin{equation}
\label{eq48}
8\pi \rho  = \frac{3\ell_{1}}{(t + t_{3})^{2}} - \frac{3\alpha^{2}}
{K_{4}(t + t_{3})^{2\ell_{1}}}.
\end{equation}

When $\xi_{0} = 0$, the pressure and energy density are same as that of given in Eq.
(\ref{eq44}) of paper Camci et al. \cite{ref30}. In this case, the weak and strong 
energy conditions (\ref{eq35}) for this solution are identically satisfied when 
$\ell_{1}(1 - \ell_{1}) \geq 0$ i.e. $0 \leq \ell_{1} \leq 1$. This model is shear-free 
and expanding with $\theta = \frac{3\ell_{1}}{(t + t_{3})}$.  

\subsection{Case (III) : Let $B$ = $T^{n}$ ($n$ is a real number).} 
In this case Eq. (\ref{eq19}) gives
\begin{equation}
\label{eq49}
C = k_{3} T^{n} - 2\ell T^{\ell_{1}}
\end{equation}
and then from (\ref{eq13}), we obtain
\begin{equation}
\label{eq50}
A^{2} = k_{3} T^{2n} - 2\ell T^{\ell_{1} + n}
\end{equation}
Hence the metric (\ref{eq1}) takes the new form
\[
ds^{2} = \left(k_{3} T^{n} - 2\ell T^{\ell_{1}}\right)[dt^{2} - T^{n}dx^{2}] -
\]
\begin{equation}
\label{eq51}
e^{2\alpha x}\left[T^{2n}dy^{2} + \left(k_{3}T^{n} - 2\ell T^{\ell_{1}}\right)^{2}
dz^{2}\right]
\end{equation}
For four-parameters family of solution (\ref{eq51}), the physical and kinematical 
quantities are given by
\[
8\pi \bar{p}  = \biggl[- \frac{\ell^{2}}{2}(11n^{2} - 4n -4) T^{-3n} + \ell k_{3}
n(13n -10)T^{\ell_{1} + n} - 
\]
\begin{equation}
\label{eq52}
2k^{2}_{3} n(n - 1)T^{2n - 2}\biggr]\left(k_{3}T^{n} - 2\ell T^{\ell_{1}}\right)^{-3} 
+ \alpha^{2} T^{-n}\left(k_{3}T^{n} - 2\ell T^{\ell_{1}}\right)^{-1}, 
\end{equation}
\[
8\pi \rho  = \biggl[- \frac{\ell^{2}(11n^{2} - 4n - 14)}{2} T^{-3n} - 3\ell k_{3}
n(2 - n)T^{\ell_{1} + n} + 
\]
\begin{equation}
\label{eq53}
3k^{2}_{3} n^{2} T^{2n - 2}\biggr]\left(k_{3}T^{n} - 2\ell T^{\ell_{1}}\right)^{-3} - 
3\alpha^{2} T^{-n}\left(k_{3}T^{n} - 2\ell T^{\ell_{1}}\right)^{-1}. 
\end{equation}
The scale of expansion and the shear are obtained as
\begin{equation}
\label{eq54}
\theta = 3\left[\frac{\ell(n - 2)}{2} T^{-3n/2} + k_{3} n T^{n - 1}\right]
\left(k_{3} T^{n} - 2\ell T^{\ell_{1}}\right)^{-3/2},
\end{equation}
\begin{equation}
\label{eq55}
\sigma = \frac{kT^{-3n/2}}{2}\left(k_{3} T^{n} - 2\ell T^{\ell_{1}}\right)^{-3/2}.
\end{equation}
From (\ref{eq54}) and (\ref{eq55}), we get
\begin{equation}
\label{eq56}
\frac{\sigma}{\theta} = \frac{k}{6}\left[k_{3} nT^{-\ell_{1} + n} + \frac{\ell (n - 2)}
{2}\right]^{-1}.
\end{equation}
\subsubsection{Model I: ~ ~ Solution for $w = 0$}
When $w = 0$, Eq. (\ref{eq29}) reduces to $\xi = \xi_{0}$ = constant. Hence in this case 
Eq. (\ref{eq52}), with the use of (\ref{eq54}), leads to
\[
8\pi p  = 24 \xi_{0} D_{5}D_{4}^{-1} + \biggl[- \frac{\ell^{2}}{2}(11n^{2} - 4n -4) T^{-3n} + \ell k_{3}
n(13n -10)T^{\ell_{1} + n} - 
\]
\begin{equation}
\label{eq57}
2k^{2}_{3} n(n - 1)T^{2n - 2}\biggr]D_{4}^{-3} 
+ \alpha^{2} T^{-n} D^{-1}, 
\end{equation}
where
$$D_{4} = k_{3}T^{n} - 2\ell T^{\ell_{1}},$$
$$D_{5} = k_{3} n T^{n -1} + \frac{1}{2}(n - 2)\ell T^{-\frac{3n}{2}}.$$
\subsubsection{Model II: ~ ~ Solution for $w = 1$}
When $w = 1$, Eq. (\ref{eq29}) reduces to $\xi = \xi_{0} \rho$. Hence in this case 
Eq. (\ref{eq52}), with the use of (\ref{eq54}), leads to
\[
8\pi p = 3\xi_{0} \biggl[\{-\frac{1}{2}\ell^{2}(11n^{2} - 4n - 14)T^{-3n} - 3\ell k_{3} n(2 - n)T^{\ell_{1}
 + n} + 3k^{2}_{3}n^{2}T^{2n - 2}\}D^{-3}_{4} - 
\]
\[
3\alpha^{2}T^{-n}(k_{3} T^{n} - 2\ell T^{\ell_{1}})\biggr]\times D_{5}D^{-3}_{4} + \biggl[\{-\frac{1}{2}
\ell^{2}(11n^{2} - 4n - 4)T^{-3n} +
\]
\begin{equation}
\label{eq58}
\ell k_{3} n(13n - 10)T^{\ell_{1}  + n} - 2k^{2}_{3}n(n - 1)T^{2n - 2}\biggr]D^{-3}_{4}
+ \alpha^{2}T^{-n} D^{-1}_{4}
\end{equation}
From Eq. (\ref{eq53}), we observe that the energy density $\rho(t)$ is a decreasing functions of 
time. The energy density is always positive.

In subcase $\xi = 0$, metric (\ref{eq51}) with expressions $p$, $\rho$, $\theta$ 
and $\sigma$ for this model are same as that of solution (\ref{eq34}) of 
Camci {\it et al.} (2001). 

In subcase $\xi = 0$, $n = 0$, after an inverse transformation, metric (\ref{eq51})
reduces to the form
\begin{equation}
\label{eq59}
ds^{2} = dt^{2} - K_{5}(t + t_{4})^{2/3}dx^{2} - e^{2\alpha x}\left[dy^{2} + K^{2}_{5}
(t + t_{4})^{4/3}dz^{2}\right],
\end{equation}
where $t_{4}$ is an integrating constant. This model is different from the model 
(\ref{eq33}) by a change of scale. 

In subcase $k = 0$, same model as (\ref{eq34}) is obtained. \\

Further in subcase $\xi = 0$, $k_{3} = 0$, we see that the metric (\ref{eq51})
takes the form
\begin{equation}
\label{eq60}
ds^{2} = dt^{2} - 2\ell K_{6}(t + t_{5})^{2/3}\left[dx^{2} + e^{2\alpha x}
\left(b t^{m_{2}}dy^{2} + b^{-1}t^{-m_{2}} dz^{2}\right)\right],
\end{equation}
where $t_{5}$ is a constant, $bt^{m_{2}} = 2\ell K^{\frac{2 - 5n}{2 - n}}_{6}
(t + t_{5})^{\frac{2(2 - 5n)}{3(2 - n)}}$ and $K_{6} = \left[\frac{3(2 - n)}
{4\sqrt{2 \ell}}\right]^{2/3}$. This metric is only different from (\ref{eq33}) by
a change of sign. Also, in each of subcase the physical and kinematical properties
of obtained metric are same as that of Case(I). Therefore, we do not consider here them. \\

\subsection{Case (IV) : Let $B$ = $\tilde{\tau}^{n}$, where $n$ is any real number.} 
In this case Eq. (\ref{eq20}) gives
\begin{equation}
\label{eq61}
C = k_{4} \tilde{\tau}^{n}\exp{\left(\frac{k}{\ell_{1}} \tilde{\tau}^{\ell_{1}}
\right)}
\end{equation}
and then from (\ref{eq13}), we obtain
\begin{equation}
\label{eq62}
A^{2} = k_{4}\tilde{\tau}^{2n}\exp{\left(\frac{k}{\ell_{1}} \tilde{\tau}^{\ell_{1}}
\right)} 
\end{equation}
Hence the metric (\ref{eq1}) reduces to
\[
ds^{2} = \tilde{\tau}^{2n}\exp{\left(\frac{k}{\ell_{1}} \tilde{\tau}^{\ell_{1}}\right)}
\left[\tilde{\tau}^{n}\exp{\left(\frac{2k}{\ell_{1}} \tilde{\tau}^{\ell_{1}}\right)}
- dx^{2}\right]
\]
\begin{equation}
\label{eq63}
- e^{2 \alpha x}\left[dy^{2} + \exp{\left(\frac{2k}{\ell_{1}} \tilde{\tau}^{\ell_{1}}
\right)} - dz^{2}\right],
\end{equation}
where, without any loss of generality, the constant $k_{4}$ is taken equal to $1$. 
Expressions for physical and kinematical parameters for the model (\ref{eq63}) are 
not given here, but it is observed that the properties of the metric (\ref{eq63}) 
are same as that of the solution (\ref{eq38}), i.e. the Case (II). \\

\section{CONCLUDING REMARKS} 

In this paper we have described a new class exact solutions of EFES for Bianchi type V 
spacetime with a bulk viscous fluid as the source of matter. Using a generation technique 
followed by Camci et al. \cite{ref30},
it is shown that the Einstein's field equations are solvable for any arbitrary cosmic 
scale function. Starting from particular cosmic functions, new classes of spatially
homogeneous and anisotropic cosmological models have been investigated for which the
fluids are acceleration and rotation free but they do have expansion and shear. For
$\alpha = 0$ in the metric (\ref{eq1}), we obtained metrics as LRS Bianchi type I model
(Hajj-Boutros \cite{ref23}; Hajj-Boutros and Sfeila \cite{ref24}; Ram \cite{ref25}; 
Mazumder \cite{ref26}; Pradhan and Kumar \cite{ref27}). 
It is also seen that the solutions obtained by Camci et al. \cite{ref30}, Ram \cite{ref25}, 
Pradhan and Kumar \cite{ref27} are particular cases (except one) of our solutions. 
The effect of bulk viscosity is to introduce a change in the perfect fluid models. 
We observe that the conclusion of Murpfy \cite{ref16} about the absence of a big bang type of 
singularity in the finite past in models with bulk viscous fluid is, in general, not true. \\
\section*{Acknowledgements}
The authors would like to thank the Harish-Chandra Research Institute, Allahabad,
India for providing facility where this work was carried out. 
\newline
\newline


\begin{thebibliography}{000}
\bibitem {ref1} 
Roy, S. R. and Prasad, A., Gen. Rel. Grav., {\bf 26}, 939 (1994).
\bibitem {ref2}
Bali, R. and Yadav, M. K., J. R. Acad. Phys. Sci., {\bf 1}, 47 (2002). 
\bibitem {ref3} 
Farnsworth, D. L., J. Math. Phys., {\bf 8}, 2315 (1967).
\bibitem {ref4} 
Maartens, R. and Nel, S. D., Comm. Math. Phys., {\bf 59}, 273 (1978).
\bibitem {ref5} 
Wainwright, J., Ince, W. C. M. and Marshman, B. J., Gen. Rel. Grav., {\bf 10}, 259 (1979).
\bibitem {ref6} 
Hewitt, C. G. and Wainwright, J., Phys. Rev. D, {\bf 46}, 4242 (1992). 
\bibitem {ref7} 
Collins, C. B., Comm. Math. Phys., {\bf 39}, 131 (1974).
\bibitem {ref8}
Benerjee, A. and Sanyal, A. K., Gen. Relat. Grav., {\bf 20}, 103 (1988).
\bibitem {ref9} 
Meena, B. L. and Bali, R., Pramana - journal of phys., {\bf 62}, 1007 (2004).
\bibitem {ref10} 
Pradhan, A. and Rai, A., Astrophys. Space Sci. {\bf 291}, 149 (2004).
\bibitem {ref11} 
Pradhan, A., Yadav, L. and Yadav, A. K., Czech. J. Phys., {\bf 54}, 487 (2004). 
\bibitem {ref12}  
Misner, C. W., Nature, {\bf 214}, 40 (1967). \\
Misner, C. W., Astrophys. J., {\bf 151}, 431 (1968).
\bibitem {ref13}
Weinberg, S., Astrophys. J., {\bf 168}, 175 (1971).
\bibitem {ref14}
Nightingale, J. P., Astrophys. J., {\bf 185}, 105 (1973).
\bibitem {ref15} 
Heller, M. and Klimek, Z., Astrophys. Space Sci., {\bf 33}, 37 (1975).
\bibitem {ref16}
Murphy, G. L., Phys. Rev. D, {\bf 8}, 4231 (1973).
\bibitem {ref17}
Gr$\o$n, $\O$., Astrophys. Space Sci., {\bf 173}, 191 (1990).
\bibitem {ref18} 
Kalyani, D. and Singh, G. P., {\it in New Dirction in Relativity and Cosmology}, ed. 
Sabbata, V. de. and Singh, T. (Hydronic Press, U. S. A.), p. 41 (1997). 
\bibitem {ref19} 
Beesham, A., (1994). {\it Astro. Lett. Commu.} {\bf 29}, 233.
\bibitem {ref20}
Singh, C. P., Kumar, S. and Pradhan, A., Class. Quantum Grav., {\bf 24}, 455 (2007).
\bibitem {ref21}
Bali, R. and Pradhan, A., Chinese Phys. Lett., {\bf 24}, 585 (2007). (gr-qc/0611018). 
\bibitem {ref22}
Pradhan, A., Yadav, V. K. and Saste, N. N., Int. J. Mod. Phys. D, {\bf 11}, 857 (2002).\\
Pradhan, A. and Aotemshi, I., Int. J. Mod. Phys. D, {\bf 11}, 1419 (2002).\\
Pradhan, A. and Pandey, H. R., Int. J. Mod. Phys. D, {\bf 12}, 941 (2003).\\
Pradhan, A. and Pandey, O. P., Int. J. Mod. Phys. D, {\bf 12}, 1299 (2003).\\
Pradhan, A., Srivastav, S. K. and Jotania, K. R., Czech. J. Phys., {\bf 54}, 255 (2004).\\
Pradhan, A. and Singh, S. K., Int. J. Mod. Phys. D, {\bf 13}, 503 (2004). \\
Pradhan, A. and Pandey, P., Czech. J. Phys., {\bf 55}, 749 (2005).\\
Pradhan, A., Singh, P. K. and Jotania, K. R., Czech. J. Phys., {\bf 56}, 641 (2006).\\
Pradhan, A. and Pandey, P., Astrophys. Space Sci., {\bf 301}, 127 (2006).
\bibitem {ref23} 
Hajj-Boutros, J., {\it Lecture Notes in Physics , Gravitation, Geometry and Relativistic 
Physics}, Springer, Berlin, {\bf 212}, 51 (1984). \\
Hajj-Boutros, J., J. Math. Phys., {\bf 26}, 2297 (1985). \\
Hajj-Boutros, J., Class. Quant. Grav., {\bf 3}, 311 (1986). 
\bibitem {ref24} 
Hajj-Boutros, J. and  Sfeila, J., Int. J. Theor. Phys., {\bf 26}, 97 (1987). 
\bibitem {ref25} 
Ram, S., Gen. Rel. Grav., {\bf 21} 697 (1989). \\
Ram, S., Int. J. Theor. Phys., {\bf 29}, 901 (1990).   
\bibitem {ref26} 
Mazumder, A., Gen. Rel. Grav., {\bf 26}, 307 (1994). 
\bibitem {ref27} 
Pradhan, A. and Kumar, A., Int. J. Mod. Phys. D, {\bf 10}, 291 (2001). 
\bibitem {ref28}
Ellis, G. F. R. and MacCallum, M. A. H., Comm. Math. Phys., {\bf 12}, 108 (1969). 
\bibitem {ref29} 
Ryan, M. P. Jr. and Shepley, {\it Homogeneous Relativistic Cosmology}, Princeton 
University Press, Princeton (1975).
\bibitem {ref30} 
Camci, U., Yavuz, I., Baysal, H., Tarhan, I. and Yilmaz, I., Astrophys. Space Sci., {\bf 275}, 391 (2001).
\bibitem {ref31}
Pradhan, A., Yadav, A. K. and Yadav, L., Czech. J. Phys., {\bf 55}, 503 (2005).
\bibitem {ref32}
Pavon, D., Bafaluy, J. and Jou, D., Class. Quant. Grav., {\bf 8}, 347 (1991).   
\bibitem {ref33}
Maartens, R., Class. Quant. Grav., {\bf 12}, 1455 (1995).
\bibitem {ref34}
Zimdahl, W., Phys. Rev. D, {\bf 53}, 5483 (1996).
\bibitem {ref35}
Santos, N. O., Dias, R. S. and Banerjee, A., J. Math. Phys., {\bf 26}, 878 (1985).
\bibitem {ref36}
Belinskii, U. A. and Khalatnikov, I. M., Sov. Phys. JETP, {\bf 42}, 205 (1976).
\bibitem {ref37}
Einstein, A. and de Sitter, W., Proc. Nat. Acad. Sci., U. S. A. {\bf 18}, 213 (1932).
\bibitem {ref38} 
Hawking, S. W. and Ellis, G. F. R., {\it The Large Scale Structure of Space-Time} 
Cambridge University Press, Cambridge, P. 94 (1973).
\end{thebibliography}
\end{document}